\documentstyle[psfig,prb,aps]{revtex}
\begin{document}
\twocolumn[\hsize\textwidth\columnwidth\hsize\csname
@twocolumnfalse\endcsname
\title{
Emergence of Bulk CsCl Structure in (CsCl)$_n$Cs$^+$ Cluster Ions
}
\author{Andr\'es Aguado}
\address{Departamento de F\'\i sica Te\'orica,
Universidad de Valladolid, Valladolid 47011, Spain}
\maketitle
\begin{abstract}
The emergence of CsCl bulk structure in (CsCl)$_n$Cs$^+$ cluster ions is
investigated using a mixed quantum-mechanical/semiempirical theoretical
approach. We find that rhombic
dodecahedral fragments (with bulk CsCl symmetry) are more stable than rock-salt
fragments after the completion of the fifth rhombic dodecahedral atomic shell.
From this size (n=184) on, a new set of magic numbers should appear in the
experimental mass spectra. We also propose another experimental test for this
transition, which explicitely involves the electronic structure of the cluster.
Finally, we perform more detailed calculations in the size range n=31--33, where
recent experimental investigations have found indications of the presence of
rhombic dodecahedral (CsCl)$_{32}$Cs$^+$ isomers in the cluster beams.
\end{abstract}
\pacs{PACS numbers: 36.40.Ei, 36.40.Mr, 61.46.+w, 61.50.Lt}

\vskip2pc]

\section{Introduction}
A general goal of cluster physics is to study the emergence of bulk behavior
right from the molecular limit, by building clusters of increasing size and
following the size evolution of selected properties. From the
theoretical point of view, this ambitious plan has been largely impeded because
of the slow and nonmonotonic size evolution observed in many properties.
The predicted cluster structures are not simply related to the corresponding
bulk structures in 
many cases, which precludes the possibility of a meaningful extrapolation to
the bulk limit. Moreover, cluster structure is difficult to determine 
theoretically due to the huge increase in the number of isomers with
size, and
experimentally due to the small number of scatterers compared with the bulk
case. Nevertheless, recent advances involving ion mobility measurements,
\cite{Jar95,Mai97,Dug97,Lof00}
electron diffraction from trapped cluster ions,\cite{Mai99,Par00} or the use of
photoelectron spectra as a fingerprint of structure\cite{Fat96} have been
successful in elucidating the structures of several ionic and covalent clusters.

Abundance patterns obtained from the mass spectra of binary ionic clusters 
like the alkali halides and alkaline-earth oxides point towards a prompt
establishment of bulk rock-salt
symmetry.\cite{Cam81,Mar89,Twu90,Zie91} Theoretical
calculations have shown, however, that small sodium iodide and lithium halide 
clusters adopt ground state structures based on the stacking of hexagonal rings.
\cite{Agu97a,Agu97b} In the case of alkaline-earth oxide clusters, the large and 
coordination-dependent values of the oxide polarizabilities favor the formation
of structures with a large proportion of ions in surface sites, inducing a
delay in the emergence of bulk structural properties.\cite{Wil97,Agu99,Agu00}
Turning to
the alkali halides, bulk CsCl, CsBr, and CsI crystallize in the CsCl-type
structure, while both experimental mass spectra\cite{Cam81,Twu90} and
theoretical calculations\cite{Agu98} indicate that small clusters of those
materials adopt ground state structures which are fragments of a rock-salt
lattice. This implies that there has to be a structural phase
transition as the cluster size is increased.
Ion mobility measurements performed by L\"offler\cite{Lof00} suggest that
(CsCl)$_n$Cs$^+$ cluster ions with n=32 are specially 
compact, which might be explained by the presence of isomers with the shape of a
perfect three-shell rhombic dodecahedron (that is with bulk CsCl symmetry)
in the cluster beam. The electron diffraction experiments performed recently in
the group of Parks\cite{Par00} show that there is a substantial proportion of 
isomers with bulk CsCl symmetry for the same size.

In this theoretical work we analyce the above mentioned size-induced phase 
transition in (CsCl)$_n$Cs$^+$ cluster ions. We consider only those sizes that
correspond to geometrical shell closings for the CsCl-type (perfect rhombic
dodecahedra with n=32,87,184,335,552) and rock-salt (perfect cubes with
n=13,62,171,364,665) structural series. In doing so, we try to avoid any 
nonmonotonic size evolution in the calculated properties. 
In the upper part of Fig. 1 we display the relative number of atoms with a given
coordination as a function of N$^{-1/3}$, where N=2n+1 is the total number of
atoms in the cluster. 
In the lower part we show the number of atoms with nonbulk coordination 
relative to the total number of surface atoms. For the
largest sizes considered the proportion of bulklike atoms is dominant,
and within the surface the proportion of face-like atoms is already much larger
than those of edge and vertex-like atoms. From those sizes to the bulk, the
only meaningful size evolution of these proportions will be a slow approach to
zero of the face-like atoms. We thus expect to capture all the
physical information relevant to the phase transition by studying this set of
clusters and the corresponding bulk phases, which have been studied both
with the same theoretical model. In this way
inaccuracies related to the use of different methodologies are avoided and a
meaningful extrapolation to the bulk limit can be done.\cite{Fra95}
In a second part of the work, we explicitely analyze the structures adopted by
(CsCl)$_n$Cs$^+$ cluster ions in the size range n=31--33, in order to explain
the experimental findings of Refs. \onlinecite{Lof00} and \onlinecite{Par00}.

The rest of the paper is organized as follows: Section II includes just a brief
description of the theoretical model employed in the calculations, as a full
account of it has been given in previous publications.\cite{Agu97a,Agu99,Agu00}
In Section III we present and discuss the results of the calculations, and
Section IV summarizes the main conclusions.

\section{Theoretical model}
Cluster energies have been obtained by performing
Perturbed Ion (PI) plus polarization calculations. This is
a well tested method that describes accurately both bulk\cite{Mar97} and
cluster\cite{Agu99,Agu00} limits. 
Its theoretical foundation lies in the theory of electronic 
separability.\cite{Huz71,McW94,Fra92} Very briefly, the cluster wave
function is broken into local group functions (ionic in nature in our
case) that are optimised in a stepwise procedure. 
In each iteration, the total energy is minimized with respect to
variations of the electron density localized in a given ion, with the electron
densities of the other ions kept frozen. In the subsequent iterations each 
frozen ion assumes the role of nonfrozen ion. When the self-consistent process 
finishes,\cite{Agu97a} the outputs are the total
cluster energy and a set of localized wave functions, one for
each geometrically nonequivalent ion of the cluster. This procedure leads to a
linear scaling of the computational effort with cluster
size, which allows the investigation of large clusters
with an explicit inclusion of the electronic structure.
The cluster binding energy can be decomposed into ionic additive contributions
\begin{equation}
E^{bind}_{clus} = 
\sum_{R \in clus}(E_{add}^R -E_0^R),
\end{equation}
being E$_{add}^R$ the contribution of the ion R to the total cluster energy and
E$_0^R$ the energy of the ion R {\em in vacuo}. In this way
the contribution of ions with different coordinations to the
binding energy can be separately analyzed, which is particularly convenient
for our study. Each additive energy can be decomposed in turn as a
sum of deformation and interaction terms
\begin{equation}
E^{bind}_{clus} = \sum_{R \in clus} (E_{def}^R + \frac{1}{2}E_{int}^R),
\end{equation}
where E$_{def}^R$ is the self-energy of the ion R, measured relative to the
vacuum state, and E$_{int}^R$ contains electrostatic, exchange and repulsive
overlap energy terms.\cite{Agu97a,Agu99}
The polarization contribution to the cluster binding energy is not computed in
the actual version of the PI code, as it assumes (for computational simplicity)
that the electronic charge distribution of each ion in the cluster is
spherically symmetric. Thus, a polarization correction to the PI energy
is computed semiempirically as described in Refs.\onlinecite{Agu99,Agu00}. Bulk
polarizabilities are used for both Cs$^+$ and Cl$^-$ ions.\cite{Fow85}
This is a good approximation for the Cs$^+$ cations. The main effect on the
anion polarizabibities when passing from the bulk to a cluster environment is
an increase of the polarizabilities of those ions located on the cluster 
surface, due to the lower average coordination compared to the bulk. 
However, we have
checked that our main conclusions are not affected by an increase in the surface
chloride polarizabilities as large as 10--20 \%, which are typical values for
halides.\cite{Wil97} The short-range induction damping parameters
have been obtained through the scaling procedure validated in 
Ref.\onlinecite{Jem99}. The reliability for cluster calculations of the mixed 
quantum-mechanical/semiempirical energy model thus obtained has been checked
and shown to be high in previous publications.\cite{Agu99,Agu00}

\section{Results}
\subsection{The rock-salt to CsCl-type structural transition}
Fig. 2 shows the size evolution of the binding energy per ion.
First of all, we note that the
PI model properly reproduces the stability trend in the bulk, predicting the
CsCl structure as the most stable one. This is a tough problem for
semiempirical methods, as Pyper\cite{Pyp94} has shown that a full account of the
coordination number dependence of the self-energy and overlap contributions
is necessary to obtain the correct ground state structure. The
values of the binding energy, plotted as a function of N$^{-1/3}$, lie neatly
on a straight line. The regression coefficients obtained from a fit are
0.9998 in all cases if we exclude from the fitting the NaCl-type cluster with 
n=13, which is the smallest one. We have calculated {\em after} the fitting
procedure the energy of the 5$\times$5$\times$7 cuboid (also included in Fig. 
2), and checked that it lies on the fitted NaCl-type energy curve. This shows
that a consideration of perfect cubes (or cuboids) on one hand,
and rhombic dodecahedra on the other hand
removes the nonmonotonic behavior from the size evolution of the binding 
energies. Our results predict that the rhombic
dodecahedra become definitely more stable after the completion of the fifth 
shell of atoms, that is for n=184. The four-shell rhombic dodecahedron and the
5$\times$5$\times$7 cuboid are essentially degenerate, so both of them will
contribute to the enhanced abundance observed experimentally for n=87.
\cite{Twu90} We have not found any experimental mass spectrum
for values of n as high as 184, but we predict
that a new set of magic numbers, corresponding to the closing of rhombic
dodecahedral atomic shells, should emerge from this size on. The magic numbers
corresponding to the closing of perfect cubic shells will probably not
disappear still at that specific size from the mass spectra, 
because they do not coincide with the 
CsCl shell closings, and complete cubes can remain more
stable than incomplete rhombic dodecahedra until larger values of n are reached.
Polarization has little influence on these general results, and only affects
the energetic ordering of the two essentially degenerate isomers mentioned
above.

Now we turn to an analysis of the physical factors responsible for this
transition. In Fig. 3 we show the binding
energy per ion, averaged over subsets of ions with a fixed coordination.
The contribution of bulklike ions to the binding energy favors always the
formation of CsCl-type structures.
However, the contribution of facelike ions
favors the formation of rock-salt fragments.
As soon as the proportion of bulk ions is larger
than that of surface atoms, which occurs after the completion of the fifth
rhombic dodecahedral atomic shell, fragments of the CsCl-type lattice become
more stable. The energy contribution of those ions in edge positions is
approximately the same for both structural families except for the smallest
clusters; finally, corner atoms
favor the CsCl-type structures, but their small relative number results in a
very small contribution to the total energy for those sizes where the 
transition occurs.

Fig. 3 has reduced the structural phase transition in
(CsCl)$_n$Cs$^+$ cluster ions to an essentially bulk 
effect. By this we mean that CsCl-type structures become more stable as soon as
the proportion of bulklike atoms is dominant.
To complete our discussion we have then to address the stability 
question in the bulk. This is more easily understood by analyzing the
reasons why other alkali halides like NaCl or CsF do not crystallize in the
CsCl-type lattice. The largest contribution to the binding energy of an ionic
crystal is the Madelung energy term E$_M=\frac{A_M}{R_e}$, with $A_M$
the Madelung constant and $R_e$ the equilibrium interionic distance. The
Madelung constant of the CsCl-type lattice (1.762675) is larger than that of
the rock-salt lattice (1.747565), so were the value of R$_e$ the same for both
structures, the CsCl-type would always be more stable.
We have solved for the electronic structure of NaCl and CsF
crystals in the CsCl-type structure at a nonequilibrium value of the interionic
distance, chosen in such a way that the Madelung energy term is exactly
the same as in
the corresponding rock-salt lattice at equilibrium. In the case of
NaCl, E$_{add}$(Na$^+$) favors the CsCl-type structure, but 
E$_{add}$(Cl$^-$) largely favors the rock-salt phase. The main reason is the
large anion-anion overlap at that artificial distance, that is the Na$^+$
cation is so small compared to the Cl$^-$ anion that eight anions can not
be packed efficiently around a cation. In CsF the situation is
reversed, and it is the cation-cation overlap that is too large.
This demonstrates that the stability situation in the bulk is a purely
packing effect: in CsCl, CsBr and CsI, the large value of the
cation-anion size ratio allows for an equilibrium
interionic approach in the CsCl-type structure close enough as to obtain a
Madelung energy term more negative than in the rock-salt phase, without a 
large overlap interaction between like ions. 
The same is true for the bulklike ions in the clusters studied, and so when
those ions begin to dominate the energetics, the bulklike fragments become more
stable.

We have made a prediction above that can be tested experimentally, namely the
emergence of a new set of magic numbers from n=184 on. Here we propose another,
perhaps more indirect, experimental test. In Fig. 4, the eigenvalues of the 3p
orbitals of Cl$^-$ (with opposite sign) are plotted as a function of N$^{-1/3}$.
We have a band of eigenvalues for each size because the anions occupy 
nonequivalent positions in the clusters. As the clusters under study are formed
by closed shell ions whose wave functions are strongly localized, it can be
assumed that an electron is extracted from a specific localized orbital when the
cluster is ionized. This is the lowest bound 3p orbital, which corresponds
always to a chloride anion with a low coordination. Thus the dashed lines
represent the size evolution of the vertical ionization potential IP (in the
Koopmans' approximation) for both structural families. For the rock-salt
series, that size evolution is approximately linear in N$^{-1/3}$, but for the
CsCl-type series it shows a more or less oscillating behavior, which should be
detected in experimental measurements of the vertical IP if rhombic
dodecahedra actually are the ground state structures from a given size on. 
We can explain these different electronic behaviors in a very simple way: 
in the rock-salt clusters the eight corner sites are always occupied by Cs$^+$
cations. The weakest bound electron corresponds always to a Cl$^-$ anion with
fourfold coordination, namely anyone of those closer to the corner cation sites.
On the other hand, rhombic dodecahedra have fourteen corner sites. When the
number of atomic shells is even, all these sites are occupied by Cs$^+$
cations, but when that number is odd, eight of them are cationic sites and the
other six anionic sites. Thus the nonmonotonic behavior of the vertical IP is
due to the different local coordination of the Cl$^-$ anion to which the
weakest bound electron is attached as the number of atomic shells increases.

\subsection{Structures of (CsCl)$_n$Cs$^+$ (n=31--33) and comparison to 
            experiment}
We finish our study with an explicit consideration of (CsCl)$_n$Cs$^+$
clusters in the size range n=31--33, the range covered in the experiments of
L\"offler\cite{Lof00} and Parks.\cite{Par00} Specifically, we have considered
the most compact 7$\times$3$\times$3, 4$\times$4$\times$4 and
5$\times$4$\times$3 rock-salt 
structures, and the three-shell rhombic dodecahedron, with some atoms added or
removed from different positions. The binding energies are shown in Table I.
The ground state (GS) structure of
(CsCl)$_{31}$Cs$^+$ is a complete 7$\times$3$\times$3 cuboid. The 
4$\times$4$\times$4 fragment with an anion removed from a corner position is
slightly less stable, and the lowest energy rhombic dodecahedral isomer we
have obtained has a still lower stability. For n=32, the
complete three-shell rhombic dodecahedron becomes the GS isomer. All the
different incomplete rock-salt fragments have a smaller binding energy. For
n=33, the different rock-salt isomers are essentially degenerate, but the
CsCl-type structure is found again at a higher energy. This sequence of GS
structures for (CsCl)$_n$Cs$^+$ clusters is consistent with the experimental
findings.\cite{Lof00,Par00}
The relative mobility is a local maximum for n=32, as the perfect
three-shell rhombic dodecahedron is evidently more compact than the complete
7$\times$3$\times$3 (CsCl)$_{31}$Cs$^+$ cuboid or any of the incomplete 
rock-salt structures obtained for n=33. Also, the energetical ordering of the
isomers is consistent with the large proportion of CsCl-type isomers found
for n=32 in the electron diffraction experiments.

\section{Summary}

We have reported a computational study of the size-induced rock-salt to 
CsCl-type structural phase transition in (CsCl)$_n$Cs$^+$ cluster ions. For
this purpose, the Perturbed Ion (PI) method, supplemented with a semiempirical
account of polarization effects, has been employed. Only cluster ions with an
atomic closed-shell configuration have been considered in order to avoid
nonmonotonic behavior in the calculated properties. Moreover, we have employed
the same theoretical model to study both cluster and bulk limits, which allows
for a meaningful extrapolation strategy. The main result is that rhombic 
dodecahedral isomers become definetely more stable than rock-salt structures
after the completion of the fifth rhombic-dodecahedral atomic shell, that is
for a size n=184. Thus, it is predicted that a new set of magic numbers, 
reflecting the establishment of the new structural symmetry, should emerge from
that size on. The size evolution of the vertical ionization potential of the
cluster ions should also be a good experimental fingerprint of the transition.
In order to explain the nature of the transition, an analysis of the binding
energy into ionic components has been performed. The result is quite simple:
bulklike ions always prefer to have a CsCl-type environment, even for the
smallest cluster sizes (this has been shown to be a purely packing effect), 
while surface-like atoms prefer to adopt rock-salt
structures. The transition occurs as soon as the proportion of bulklike atoms
is large enough to dominate the energetics of the whole cluster. One of the
possibilities advanced by Parks consistent with his experimental results
\cite{Par00} is the existence of isomers with mixed symmetry. Our results
indicate that the formation of isomers with a CsCl-type core and a 
rock-salt-type surface could be energetically favored, if the strain accumulated
in the bonds at the interface region separating both phases can be conveniently 
relaxed. This point
deserves further investigation.

The structures adopted by (CsCl)$_n$Cs$^+$ cluster ions have been more carefully
studied in the size range n=31--33, which has been covered in the experimental
investigations. Our results are consistent with the experimental findings, and
show that the three-shell rhombic dodecahedron is the lowest energy isomer for 
n=32.

{\bf ACKNOWLEDGMENTS}\\
This work has been supported by DGES 
(Grant PB98-0368)
and Junta de Castilla y Le\'on (VA70/99).
The author is indebted to J. M. L\'opez for a careful reading of the manuscript.

\newpage

{\bf Captions of Tables.}

{\bf Table I} Binding energy, in eV/ion, of the different rock-salt and
              CsCl-type structures for the size range n=31--33.

{\bf Captions of Figures.}

{\bf Figure 1} Size evolution of the number of atoms with a given coordination,
relative to the total number of atoms (upper half) or to the total number of
surface atoms (lower half). The left half refers to CsCl-type symmetry and
the right half to rock-salt symmetry.

{\bf Figure 2} Size evolution of the binding energy per ion for both CsCl-type
and rock-salt structural families, with (lower half) and without (upper half)
the inclusion of polarization corrections. The value of N$^{-1/3}$ at the
transition point has been indicated with an arrow.

{\bf Figure 3} Size evolution of the binding energy contributions from ions
with different coordinations. Full circles represent ions in the CsCl-type
structures and squares represent ions in the rock-salt structures.

{\bf Figure 4} Size evolution of the 3p orbital eigenvalues of chloride anions.
The dashed line represents the variation of the vertical ionization potential
in the Koopmans' approximation with size.

\onecolumn[\hsize\textwidth\columnwidth\hsize\csname
@onecolumnfalse\endcsname

\begin {table}
\begin {center}
\begin {tabular} {|c|c|c|c|c|c|}
n=31 & & n=32 & & n=33 & \\
\hline
 Structure & Energy (eV/ion) & Structure & Energy (eV/ion) & Structure & Energy
(eV/ion) \\
\hline
7$\times$3$\times$3 & 3.032 & 7$\times$3$\times$3+2 & 3.021 & 7$\times$3$\times$3+4 & 3.025 \\
4$\times$4$\times$4-1 & 3.026 & 4$\times$4$\times$4+1 & 3.032 & 4$\times$4$\times$4+3 & 3.027 \\
5$\times$4$\times$3+3 & 3.019 & 5$\times$4$\times$3+5 & 2.998 & 5$\times$4$\times$3+7 & 3.028 \\
CsCl-type - 2 & 2.983 & CsCl-type & 3.048 & CsCl-type+2 & 2.980 
\end {tabular}
\end {center}
\end {table}

\begin{figure}
\psfig{figure=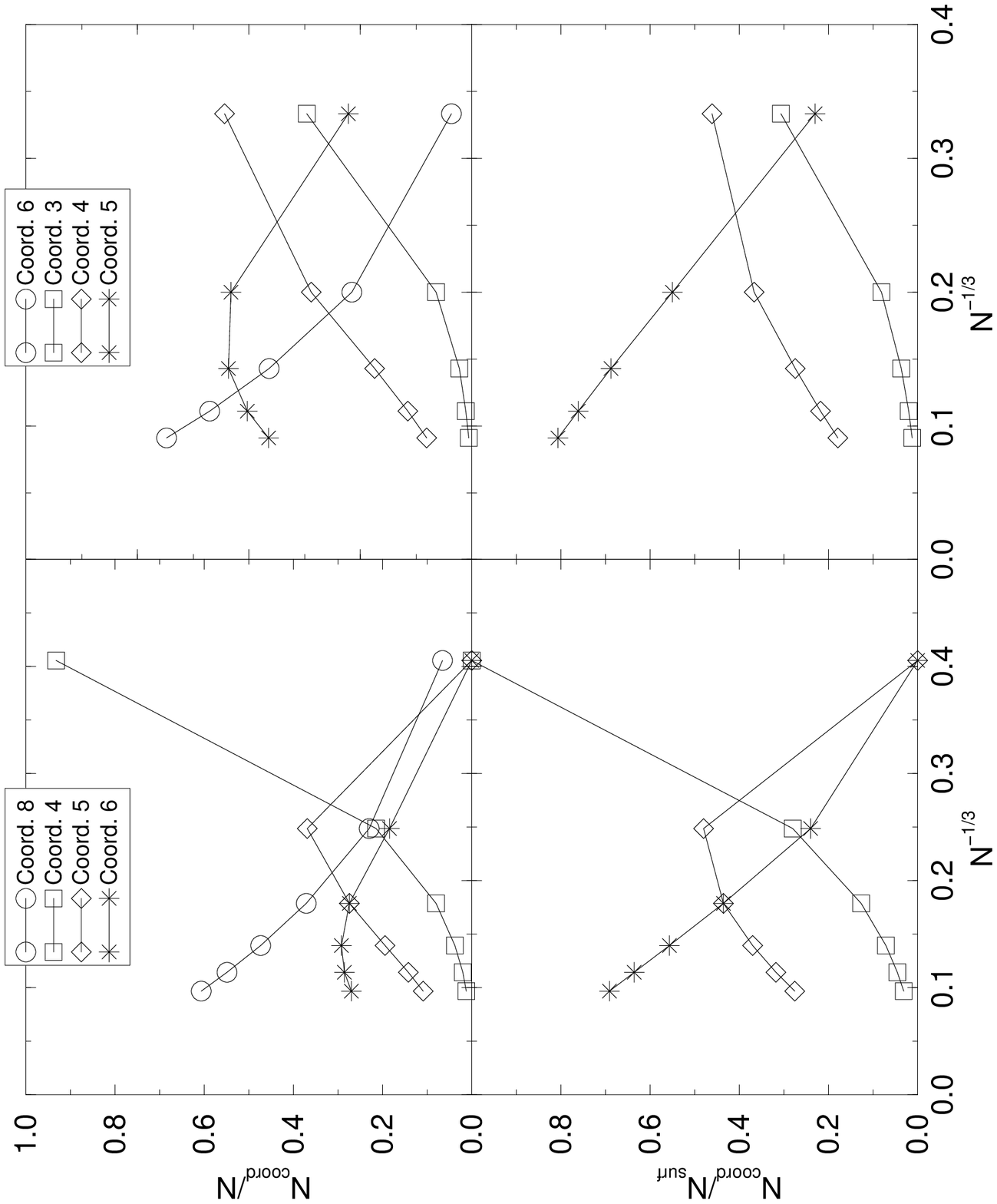}
\end{figure}

\begin{figure}
\psfig{figure=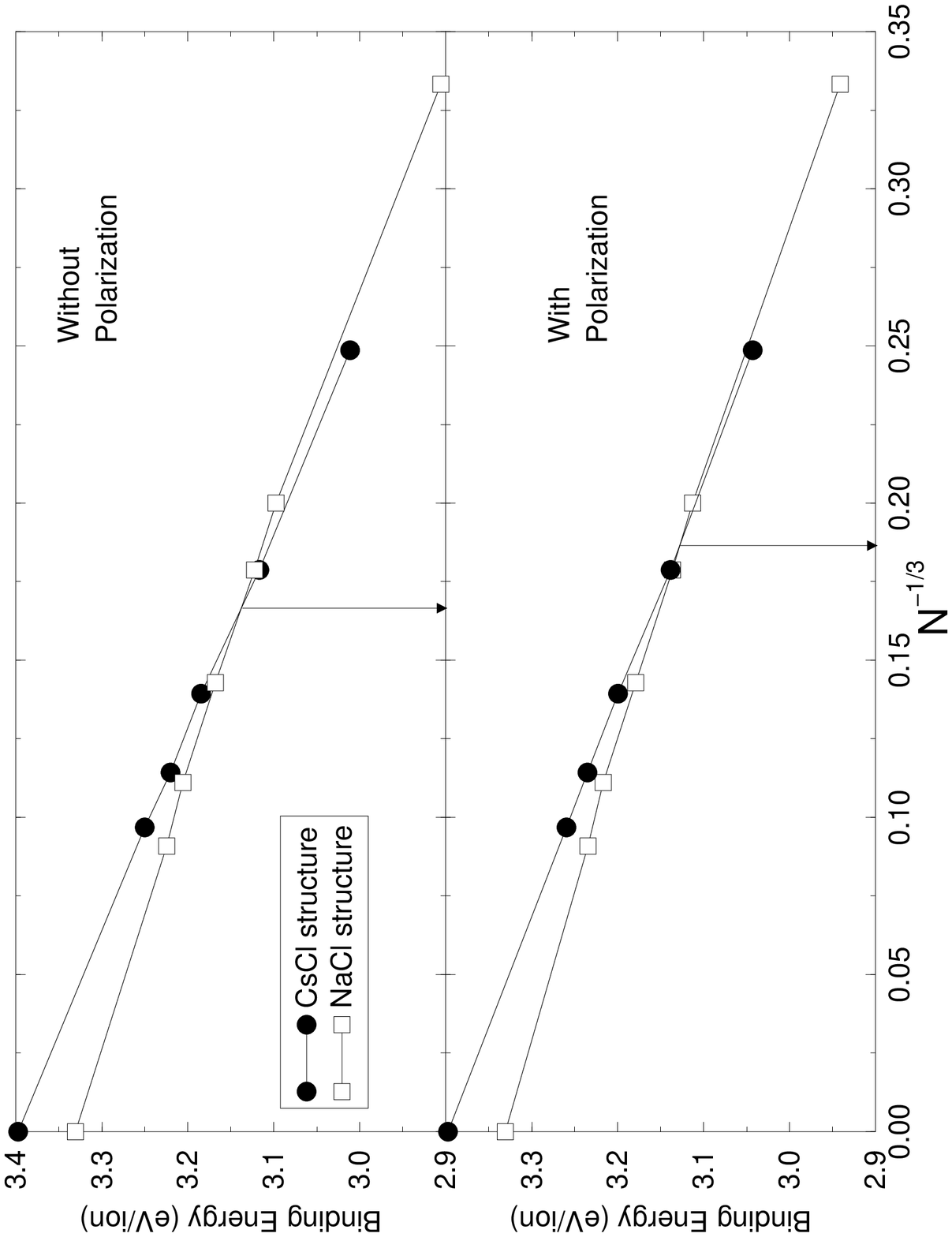}
\end{figure}

\begin{figure}
\psfig{figure=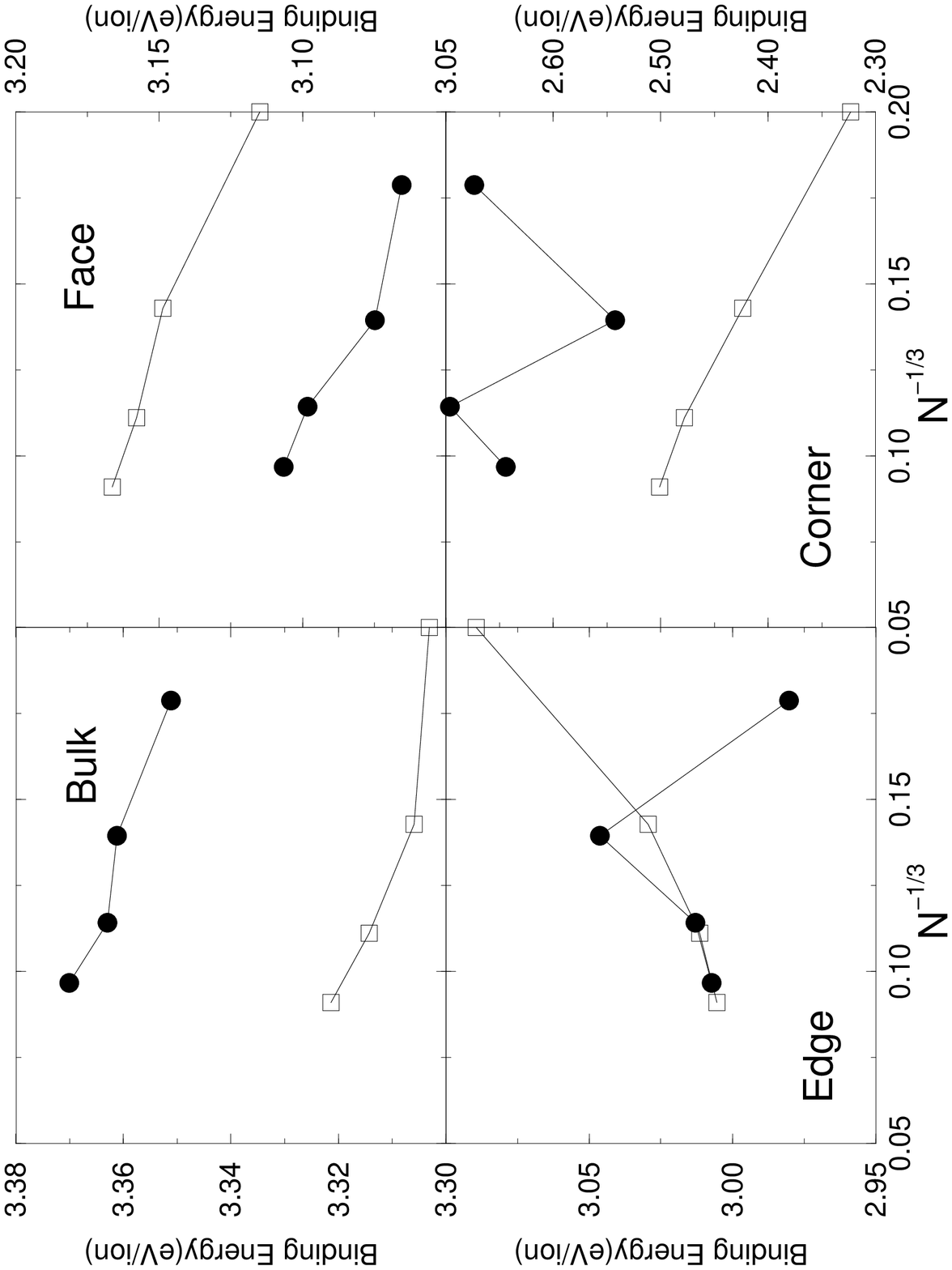}
\end{figure}

\begin{figure}
\psfig{figure=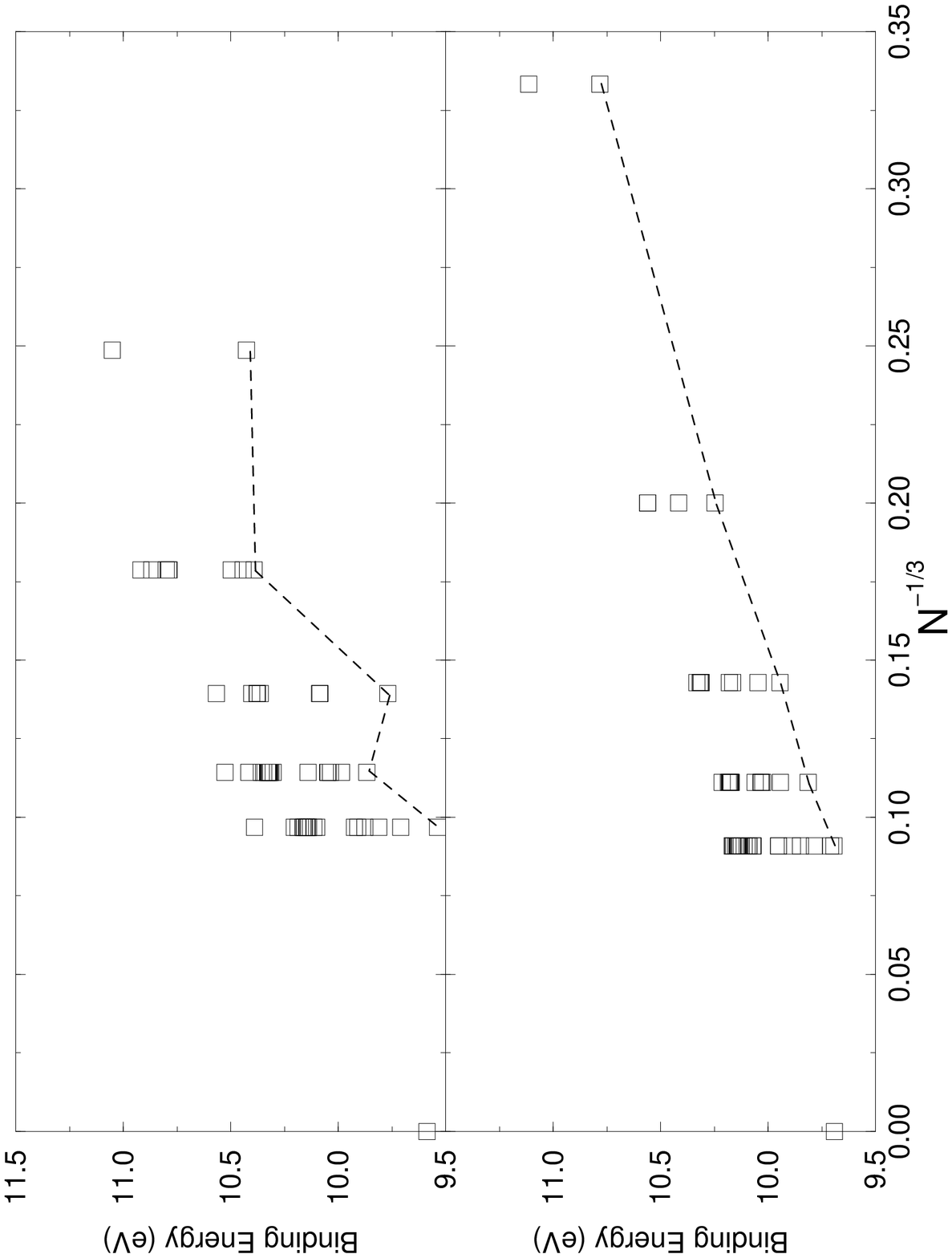}
\end{figure}

\end{document}